# A solution of the quantum time of arrival problem via mathematical probability theory

Maik Reddiger

Center for Research, Transfer, and Entrepreneurship, Anhalt University of Applied Sciences, Köthen (Anhalt), Germany

**ABSTRACT**

Time of arrival refers to the time a particle takes after emission to impinge upon a suitably idealised detector surface. Within quantum theory, no generally accepted solution exists so far for the corresponding probability distribution of arrival times. In this work we derive a general solution for a single body without spin impacting on a so called ideal detector in the absence of any other forces or obstacles. A solution of the so called screen problem for this case is also given. After discussing the shortcomings of the so called 'absorbing boundary condition', which is arguably the natural approach within quantum mechanics, we construct the ideal detector model via mathematical probability theory. This detector model assures that the probability flux through the detector surface is always positive, so that the corresponding distributions can be derived via an approach originally suggested by Daumer, Dürr, Goldstein, and Zanghì. The resulting dynamical model is based on an adaption of the Madelung equations and is, strictly speaking, not compatible with quantum mechanics. Still, it is well-described within geometric quantum theory. Geometric quantum theory is a novel adaption of quantum mechanics, which makes the latter consistent with mathematical probability theory. Implications to the general theory of measurement and avenues for future research are also provided. Future mathematical work should focus on finding an appropriate distributional formulation of the evolution equations and studying the well-posedness of the corresponding Cauchy problem.

## 1. Introduction

In simple terms, the time of arrival problem concerns the question of how long it takes a quantum particle after emission to reach a suitably idealised detector surface.

The status of the problem within contemporary physics is a paradox in the following sense: on the one hand, it appears intuitively simple. On the other hand, common quantum-mechanical concepts do not seem to suffice to adequately address it:

**CONTACT** Maik Reddiger maik.reddiger@hs-anhalt.de Center for Research, Transfer, and Entrepreneurship, Anhalt University of Applied Sciences, Hubertus 1a, Köthen (Anhalt) 06366, Germany
This article has been corrected with minor changes. These changes do not impact the academic content of the article.

famously, time is not an 'observable' in the theory (cf. Sec. A.8 in [1–3]). Accordingly, a large variety of conceptually different approaches have been suggested so far (see e.g. [4–14]) and no scientific consensus on the resolution of the problem has been attained. Reviews on the subject are given in [15] and Section 2 of [16].

Of course, the indeterminateness of quantum phenomena requires a restatement of the problem: what is asked for is the corresponding probability distribution. It is the task of quantum theory to predict this distribution for any given setup.

The purpose of this work is to provide a systematic, yet at this point not mathematically rigorous solution to the problem through the use of mathematical probability theory [17–19] and a systems of equations discovered by Madelung in 1926 [20–22]. Though the approach can in principle be generalised, we restrict ourselves to the non-relativistic regime and the case of a single body 'without spin' impacting on a so called ideal detector in the absence of any other forces or obstacles.

While the question of the (first) arrival time distribution on a non-interacting surface in space might be within the reach of scientific inquiry, this situation is not of empirical concern – with the single exception of the asymptotic case treated in quantum scattering theory [23–25]. Only the latter can be reasonably linked to experiment. It is therefore more appropriate to refer to 'times of impact' rather than 'times of arrival'. The construction of a detector model is therefore imperative [1,13], at least for the purpose of making empirical predictions 'in the near field' (cf. [23]).

Accordingly, we qualitatively define an *ideal detector* as a surface, which captures the body upon impact and holds it fixed, but does otherwise not interact with it. That in any actual experiment the body will not stay fixed on the surface shall not be of any greater concern, since for modelling purposes we are only interested in where the body impacts and when. It is a choice of convenience that the body is modelled to stay on the surface after impact. In personal conversations with experimental physicists it was confirmed that contemporary detectors are highly efficient and that such an idealised detector model may indeed be of empirical use.

An overview of this work as well as its main results is given hereafter.

Section 2 provides a refined problem statement. The problem is then discussed in the context of the so called 'absorbing boundary condition' [13,14,23]. We find that that approach is rather natural within the theory of quantum mechanics and nonetheless objectionable on physical grounds. In Section 3, we employ mathematical probability theory to motivate the introduction of a probability surface density $\sigma$ on the detector surface $\mathcal{D}$ and construct the aforementioned ideal detector model.

In Section 4, we show how this model leads to an adapted form of Madelung's equations: the *(free) 1-body Madelung equations in the presence of an ideal detector*. This is a constrained system of nonlinear partial differential evolution equations in the (volume) probability density $\rho$ in the region $\Omega$, the probability surface density $\sigma$ on the detector surface $\mathcal{D} = \partial\Omega$, as well as the probability current velocity $\vec{v}$ (see Equations (6) and (12) below for definitions in terms of a given wave function). To state the equations, we recall the definition of the stochastic velocity [26–28]:

$$\vec{u} = \frac{\hbar}{2m}\frac{\nabla\rho}{\rho}. \tag{1}$$

We further denote the outward-pointing unit normal vector field on $\mathcal{D}$ by $\vec{n}$, the Heaviside step function by $\theta$, the Dirac distribution on $\mathcal{D}$ by $\delta_{\mathcal{D}}$ and the restriction of a field to $\mathcal{D}$ by $\upharpoonright_{\mathcal{D}}$. The system of equations is then given by

$$m\left(\frac{\partial \vec{v}}{\partial t} + (\vec{v}\cdot\nabla)\vec{v}\right) = m(\vec{u}\cdot\nabla)\vec{u} + \frac{\hbar}{2}\Delta\vec{u} \tag{2a}$$

$$\frac{\partial \rho}{\partial t} + \nabla\cdot\left(\rho\,\vec{v}\right) = \vec{n}\cdot\left(\rho\,\vec{v}\right)\upharpoonright_{\mathcal{D}}\ \theta\big(-\vec{n}\cdot\left(\rho\,\vec{v}\right)\upharpoonright_{\mathcal{D}}\big)\ \delta_{\mathcal{D}} \tag{2b}$$

$$\frac{\partial \sigma}{\partial t} = \vec{n}\cdot\left(\rho\,\vec{v}\right)\upharpoonright_{\mathcal{D}}\ \theta\big(\vec{n}\cdot\left(\rho\,\vec{v}\right)\upharpoonright_{\mathcal{D}}\big), \tag{2c}$$

subject to the constraints

$$\nabla\times\vec{v} = 0 \tag{2d}$$

$$\int_{\Omega}\rho\,\mathrm{d}^3 r + \int_{\mathcal{D}}\sigma\,\mathrm{d}S = 1. \tag{2e}$$

We also derive a formal Schrödinger equation, Equation (33) below. A proposal for a corresponding weak formulation is given in the appendix.

This dynamical part of the solution to the time of arrival problem is heuristic in the sense that at this point Equation (2a) are only formal. That is, a rigorous solution requires the formulation of a corresponding system of distributional PDEs and a well-posedness result under suitable conditions on the initial data (cf. [29]). Since the aim of this work is to focus on the main ideas and such mathematical questions are beyond its scope, at this point the proposed solution only meets the standards of theoretical physics in terms of rigour, not those of mathematical physics.

In Section 5, we derive the impact time distribution $\mathbb{T}$ in terms of the normal component of the (time-dependent) probability current density

$$\vec{j} = \rho\,\vec{v} \tag{3}$$

with respect to $\mathcal{D}$. The probability that the body impinges on $\mathcal{D}$ between time $t \geq 0$ and $t + \Delta t$ for $\Delta t > 0$ is given by

$$\mathbb{T}([t, t+\Delta t]) = \int_t^{t+\Delta t}\mathrm{d}t'\int_{\mathcal{D}}\mathrm{d}S\ \vec{n}\cdot\vec{j}_{t'}\upharpoonright_{\mathcal{D}}\ \theta\big(\vec{n}\cdot\vec{j}_{t'}\upharpoonright_{\mathcal{D}}\big). \tag{4}$$

From this expression, a solution of the so called 'screen problem' [23,30] is also inferred.

Section 6 concludes with a discussion of the model in the context of the theory of measurement and the encompassing theoretical framework, as well as potential avenues for further research.

## 2. Problem description

We consider a connected, closed subset $\Omega$ of $\mathbb{R}^3$ such that its interior $\Omega$ is nonempty and its topological boundary $\partial\Omega$ is a smooth surface. We interpret the latter as an ideal detector surface $\mathcal{D} = \partial\Omega$. For instance, $\Omega$ may be the half-space $(0, \infty)\times\mathbb{R}^2$ or a closed ball of some given radius.

The initial position probability density $\rho_0 \in L^1(\Omega, \mathbb{R})$ of the body we require to be compactly supported in $\Omega$. Furthermore, the support of $\rho_0$, denoted by $\operatorname{supp}\rho_0$, ought to be connected (cf. Remark 3.4 in [29]). Figure 1 provides a graphical depiction. Physically, the body is emitted from a particle gun at time $t = 0$ within the region $\operatorname{supp}\rho_0$ in $\Omega$. The body then evolves freely, up until it impinges on the detector $\mathcal{D}$. The function $\rho : t \longmapsto \rho_t$ describes the time evolution of that position probability density in space.

A more refined physical interpretation of $\rho$ may be obtained on the basis of the concept of a statistical ensemble [31–33]. For an elaboration thereof in the context of this work, see Section 5.1 in [34], Section 2.1 in [35], as well as Section 3 in [28]).

The choice of the detector model is crucial for the physical description, since it has a direct impact on the evolution of $\rho : t \longmapsto \rho_t$ (see also [1]). Corresponding evolution laws for the whole system need to be determined before one can address more advanced physical questions such as arrival time distributions. Proposals for the latter, which do not account for this point, may serve as approximations in some instances, but they do not fully solve the problem. Hence, we seek to implement the ideal detector model in terms of dynamical equations for the body.

A seemingly natural approach was given by Werner in the 1980s [14] (see also [36]). Therein the detector model is motivated from abstract considerations within the theory of quantum mechanics. It is not based on given physical properties of the detector.

While this work advocates for a different approach, it is nonetheless illustrative to see explicitly how the two approaches differ and why.

To recapitulate the results of [14], we consider a normalised, scalar wave function $\Psi_0$ on $\Omega$. For $t \geq 0$ we let it evolve according to the free Schrödinger equation

$$i\hbar \frac{\partial}{\partial t} \Psi_t = -\frac{\hbar^2}{2m} \Delta \Psi_t. \tag{5}$$

The Born rule for position now states that the probability density for the position of the body is

$$\rho = |\Psi|^2. \tag{6}$$

The aforementioned physical condition is then implemented by imposing a suitable boundary condition on $\Psi_t$ at $\mathcal{D}$ for every $t \geq 0$ (cf. [13,14]).

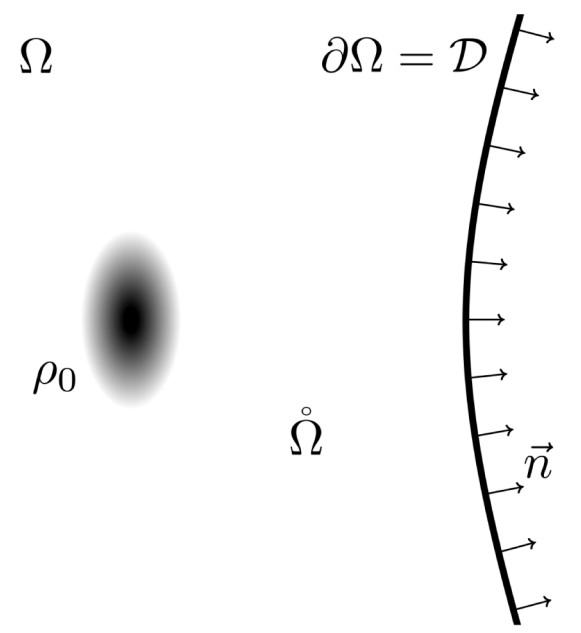

**Figure 1.** Sketch of the physical setup. An initial probability density $\rho_0$ is placed in a region $\Omega \subseteq \mathbb{R}^3$ with interior $\mathring{\Omega}$ in front of a detector screen $\partial\Omega = \mathcal{D}$ with unit normal vector field $\vec{n}$. The task is to determine the time evolution of the density $\rho$ and predict the impact time distribution on $\mathcal{D}$.

Denote by $\vec{n}$ the unit normal vector field of $\mathcal{D}$ pointing away from $\Omega$, denote restrictions of functions in $\Omega$ to $\mathcal{D}$ (if defined) by $\restriction_{\mathcal{D}}$, and let $\beta$ be some (suitably regular) complex-valued function on $\mathcal{D}$. Then the boundary condition is

$$\vec{n} \cdot (\nabla\Psi) \restriction_{\mathcal{D}} = \beta \Psi \restriction_{\mathcal{D}} \tag{7}$$

with $\mathrm{Im}(\beta) \geq 0$ and a further technical requirement on $\beta$, which is only of minor interest here (cf. Secs. II and IV in [14]). For constant $\beta$ this constitutes a homogeneous Robin boundary condition [37,38].

Again, if we argue on the basis of quantum-mechanical concepts, Equation (5) in conjunction with the condition (7) is rather natural. First, it does not modify the basic dynamical equation, Equation (5). Second, condition (7) may be formulated purely in terms of the wave function $\Psi$ and its derivatives. And third, condition (7) is linear in $\Psi$.

Still, Cavendish and Das [23] argue that for constant $\beta$ this approach yields qualitatively incorrect predictions for a variety of setups. The case of constant $\beta$ appears to be the one most studied [13,23,36,39–44].

To continue the analysis of Equation (7), we consider the probability current density vector field

$$\vec{j} = \frac{\hbar}{m} \mathrm{Im}\left(\Psi^* \nabla \Psi\right) \tag{8}$$

(cf. Equation (3) above). Multiple authors have acknowledged its crucial role in the evolution problem [9,13,45–47].

We shall employ $\vec{j}$ to show that the boundary condition (7) indeed assures that $\mathcal{D}$ is 'absorbing' in a natural sense.

From Equation (7) and the definition of $\vec{j}$, Equation (8), it directly follows that

$$\vec{n} \cdot \vec{j} \restriction_{\mathcal{D}} = \frac{\hbar}{m} \mathrm{Im}\beta \; \rho \restriction_{\mathcal{D}} . \tag{9}$$

Since $\mathrm{Im}(\beta) \geq 0$, the above expression is positive. In the next step, we consider the implications of Equation (9) for the evolution of the total probability contained in $\Omega$. By multiplying Equation (5) with $\Psi^*$ and taking the imaginary part, we derive the so called *continuity equation*:

$$\frac{\partial \rho}{\partial t} + \nabla \cdot \vec{j} = 0. \tag{10}$$

Denote now the surface area element on $\mathcal{D}$ by dS. The rate of change of probability contained in $\Omega$, as implied by Equation (10), is then given as follows:

$$\frac{\mathrm{d}}{\mathrm{d}t} \int_\Omega \rho_t \, \mathrm{d}^3 r = - \int_\Omega \nabla \cdot \vec{j}_t \, \mathrm{d}^3 r = - \int_{\mathcal{D}} \vec{n} \cdot \vec{j}_t \restriction_{\mathcal{D}} \mathrm{dS} = - \frac{\hbar}{m} \int_{\mathcal{D}} \mathrm{Im}\beta \; \rho_t \restriction_{\mathcal{D}} \mathrm{dS}. \tag{11}$$

This rate is negative. Accordingly and as claimed, the boundary condition (7) indeed assures that probability flows out of $\Omega$ and into $\mathcal{D}$. In this sense, the detector is absorbing.

Still, there are three general reasons to hold Equation (9) as physically unsatisfactory, irrespective of the results of [23].

First, it is arguably ad hoc. Surely, in [14] it was motivated from more general quantum-mechanical considerations. Nevertheless, the condition introduces the complex function $\beta$ on $\mathcal{D}$ without specifying how it is physically determined. In order to solve the evolution problem, a particular choice of $\beta$ is required.

For the second objection, we recall the definition of $\vec{v}$. This was implicitly given in terms of $\vec{j}$ via Equations (3) and (8):

$$\rho\,\vec{v} = \frac{\hbar}{m}\mathrm{Im}\big(\Psi^*\nabla\Psi\big). \tag{12}$$

Through a fluid dynamical analogy, $\vec{v}$ may be understood as the velocity at which an 'infinitesimal' amount of probability flows at the respective point in space and time (cf. Section 5.1 in [34]). The boundary condition (7) then implies that

$$\vec{n}\cdot\vec{v}\restriction_{\mathcal{D}} = \frac{\hbar}{m}\mathrm{Im}\beta \geq 0. \tag{13}$$

Accordingly, Equation (7) fixates the normal component of $\vec{v}$ on $\mathcal{D}$ to a priori given values. Yet such a condition is arguably unphysical: the boundary condition for an absorbing ideal detector should not fixate (components of) the probability current velocity. Instead it ought to state directly how well the detector absorbs the probability current.

The last objection is closely related to the former one. For this we recall the stochastic velocity $\vec{u}$ from Equation (1). In terms of $\Psi$ it is given via

$$\rho\,\vec{u} = \frac{\hbar}{m}\mathrm{Re}\big(\Psi^*\nabla\Psi\big). \tag{14}$$

In full analogy to Equation (13), the boundary condition (7) implies the following constraint on $\vec{u}$

$$\vec{n}\cdot\vec{u}\restriction_{\mathcal{D}} = \frac{\hbar}{m}\mathrm{Re}\beta. \tag{15}$$

Thus, Equation (7) also fixates the normal component of $\vec{u}$ on $\mathcal{D}$ to a priori given values. Again, it is physically questionable, why such a constraint ought to hold for an ideal detector.

The alternative approach in the next section reproduces the successes of the boundary condition (7), while also alleviating its drawbacks.

## 3. The ideal detector model

In this section we construct a novel ideal detector model on the basis of mathematical probability theory and physical requirements imposed on the detector.

We recall that in Equation (11) we related the change of probability in $\Omega$ to the amount of probability flowing into the detector. In the approach pursued in [14], the boundary condition (7) at $\mathcal{D}$ assured that the relevant quantity $\vec{n}\cdot\vec{j}\restriction_{\mathcal{D}}$ was positive, so that this probability could only enter the detector and not leave. In the literature, this is sometimes and incorrectly [48,49] called the 'no (quantum) backflow condition' and it is the origin of many of the conceptual problems arising in the context of the time of arrival problem [11,16,46,50]. Here we also aim to impose dynamical equations, which assure that the detector is absorbing in this sense.

Mathematical probability theory [17] shall be employed as a heuristic tool for constructing the ideal detector model. Mathematical probability theory was axiomatized by Kolmogorov in 1933 [18,19]. The theory in conjunction with its descendant, mathematical statistics [51], has enjoyed enormous success across the empirical sciences – so far with the exception of (fundamental) quantum physics. In a prior work [28] it was argued that the historical development of quantum theory – rather than empirical evidence – is plausibly the reason for this state of affairs (cf. Section 2 of [28] in particular). Mathematical probability theory is a standard topic in the mathematical literature. An introduction may be found, for instance, in [17].

In mathematical probability theory the notion of a probability space is fundamental (cf. Def. 1.38(iii) and Section 1.5 in [17]). By definition, a *probability space* is a tuple $(\Omega, \mathcal{A}, \mathbb{P})$ such that $\Omega$ is a nonempty set, $\mathcal{A}$ is a suitably chosen $\sigma$-algebra on $\Omega$ and $\mathbb{P}: \mathcal{A} \to [0, 1]$ is a measure with the property that $\mathbb{P}(\Omega) = 1$. $\Omega$ is called the *sample space*. $\mathcal{A}$ may be viewed as the collection of all (random) events. Hence, a *(random) event* is a subset $A$ of $\Omega$, which is also an element of $\mathcal{A}$. The number $\mathbb{P}(A)$ is then the *probability of* A. The measure $\mathbb{P}$ is called a *probability measure* or a *probability distribution* (if it is associated to a random variable, see Section 1.5 in [17]).

In quantum theory, the probability measure is usually time-dependent. We write $\mathbb{P}_t$ for the probability measure at time $t \in \mathbb{R}$ (cf. Section 4 in [28]). Accordingly, the main goal of this section is to construct a suitable time-dependent probability space for the problem at hand.

At any given time $t$, the body is either in the space $\Omega$ or on the detector $\mathcal{D}$. The (volume) probability density $\rho$ determines the probability for finding the body in a given region of space by integrating it over that region. Yet we should also be able to compute the probability of finding the body in a given area on the detector. This motivates the introduction of a *(time-dependent) surface probability density* $\sigma$ on $\mathcal{D}$. In order to assure that $\rho$ and $\sigma$ indeed determine probabilities, we must impose that

$$1 = \int_{\Omega} \rho_t \, \mathrm{d}^3 r + \int_{\mathcal{D}} \sigma_t \, \mathrm{dS} \tag{16}$$

holds for all $t \geq 0$.

Equation (16) indeed enables us to construct a (time-dependent) probability space on the set $\Omega$. It is given by $\Omega$ together with the Lebesgue sets $\mathcal{B}^*(\Omega)$ and the probability measure

$$\mathbb{P}_t : \mathcal{B}^*(\Omega) \to [0, 1] \, , \quad U \mapsto \mathbb{P}_t(U). \tag{17}$$

For any $U \in \mathcal{B}^*(\Omega)$ and all $t \geq 0$, the measure $\mathbb{P}_t$ is defined via

$$\mathbb{P}_t(U) = \int_{U \cap \Omega} \rho_t \, \mathrm{d}^3 r + \int_{U \cap \mathcal{D}} \sigma_t \, \mathrm{dS}. \tag{18}$$

Equation (16) is then equivalent to

$$\mathbb{P}_t(\Omega) = 1. \tag{19}$$

Hence, $\mathbb{P}_t$ is indeed a probability measure. $\mathbb{P}_t(U)$ gives the probability that the body is either in space in the region $U \cap \Omega$ or on the detector surface in the region $U \cap \mathcal{D}$.

In general and unless $\sigma_t = 0$, $\mathbb{P}_t$ is singular with respect to the Lebesgue measure on $\Omega$ (cf. Section 7.4 in [17]): for any $U \in \mathcal{B}^*(\Omega)$ the set $U \cap \mathcal{D}$ always has Lebesgue measure zero and yet the probability to find the body there may be nonzero.

From a dynamical point of view, we find that Equation (16) constitutes a time-dependent constraint on $\rho$ and $\sigma$. This constraint needs to be satisfied by the initial data. For later times it will be enforced by the dynamical equations. On the initial data we simply impose the condition: commonly, we choose $\sigma_0 = 0$, so that the body is initially situated in $\Omega$ and $\rho_0$ is taken to be a probability density in the mathematical sense of the word. For $t>0$, we satisfy Equation (16) by guaranteeing that

$$\frac{\mathrm{d}}{\mathrm{d}t}\,\mathbb{P}_t(\Omega) = \int_\Omega \frac{\partial \rho_t}{\partial t}\,\mathrm{d}^3 r + \int_\mathcal{D} \frac{\partial \sigma_t}{\partial t}\,\mathrm{d}S = 0. \tag{20}$$

If the continuity Equation (10) is assumed to hold, we may insert it into Equation (20). We then find that

$$\frac{\partial \sigma}{\partial t} = \vec{n} \cdot \vec{j}\restriction_\mathcal{D} \tag{21}$$

suffices to satisfy Equation (16) for $t > 0$, provided it is satisfied for $t = 0$. Equation (21) then implies that the time evolution of $\sigma$ is given via

$$\sigma_t = \int_0^t \vec{n} \cdot \vec{j}_{t'} \restriction_\mathcal{D} \mathrm{d}t' + \sigma_0. \tag{22}$$

We define the detector to be absorbing, if at every point $\vec{x}$ on $\mathcal{D}$ the function $t \mapsto \sigma_t(\vec{x})$ is increasing. This is consistent with our prior usage of the term in Section 2.

A natural means to obtain this property is to impose other conditions at the boundary, so that $\partial\sigma/\partial t > 0$. An example would be condition (9) from Section 2. Yet, it appears that the problems of condition (9) generalise to this context in the following sense: such a boundary condition will involve the ad hoc introduction of new functions on $\mathcal{D}$ or the ad hoc coupling of $\vec{j}$ to other quantities already given. That is, a different detector model is needed.

In the approach pursued here, we instead *modify the continuity Equation* (10). Specifically, we use a Dirac distribution $\delta_\mathcal{D}$ on $\mathcal{D}$ to introduce a source/sink term on the detector surface:

$$\frac{\partial \rho}{\partial t} + \nabla \cdot \vec{j} = R\,\delta_\mathcal{D}. \tag{23}$$

The function $R$ is defined on $\mathcal{D}$ and formally $\delta_\mathcal{D}$ is chosen such that for any sufficiently regular function $\phi : \Omega \to \mathbb{R}$ it holds that

$$\int_\Omega \phi\,\delta_\mathcal{D}\,\mathrm{d}^3 r = \int_\mathcal{D} \phi \restriction_\mathcal{D} \mathrm{d}S. \tag{24}$$

Equation (20) constitutes a necessary condition on $R$. It is satisfied, whenever

$$R = \vec{n} \cdot \vec{j} \restriction_\mathcal{D} - \frac{\partial \sigma}{\partial t}. \tag{25}$$

It follows from the definition that the detector is absorbing, if and only if the rate $\partial\sigma/\partial t$ is positive. How this is assured, depends on the particular detector model.

From Section 1 we recall the physical requirement on an ideal detector, that, once the body impinges on the detector surface $\mathcal{D}$, it stays thereon and is thus permanently removed from the interior $\Omega$. We call this property *perfect absorption*, so that ideal detectors are, by definition, *perfectly absorbing*. Mathematically, we implement this property by introducing the Heaviside function $\theta$ and simply setting

$$\frac{\partial\sigma}{\partial t} = \vec{n}\cdot\vec{j}\restriction_{\mathcal{D}}\ \theta\big(\vec{n}\cdot\vec{j}\restriction_{\mathcal{D}}\big). \tag{26}$$

Equation (22) is then replaced by

$$\sigma_t = \int_0^t \vec{n}\cdot\vec{j}_{t'}\restriction_{\mathcal{D}}\ \theta\big(\vec{n}\cdot\vec{j}_{t'}\restriction_{\mathcal{D}}\big)\mathrm{d}t' + \sigma_0. \tag{27}$$

By construction, perfectly absorbing detectors are absorbing.

Of course, if $\vec{n}\cdot\vec{j}\restriction_{\mathcal{D}}$ were to take on negative values, then it may seem like the choice of Equation (26) would amount to ignoring the probability flowing out of the detector surface. However, Equation (25) in conjunction with Equation (26) yields the following expression:

$$R = \vec{n}\cdot\vec{j}\restriction_{\mathcal{D}}\ \theta\big(-\vec{n}\cdot\vec{j}\restriction_{\mathcal{D}}\big). \tag{28}$$

We shall show that Equation (28) implies that there is no probability flowing from $\mathcal{D}$ into $\Omega$. To this end, we consider an arbitrary (sufficiently regular) $U \in \mathcal{B}^*(\Omega)$ and the corresponding (outward pointing) surface normal $\vec{n}'$ on $\partial U$ with differential surface element $\mathrm{d}S'$. By decomposing

$$\partial U = \big(\partial U \setminus \mathcal{D}\big) \cup (\partial U \cap \mathcal{D}) \tag{29}$$

and using Equation (28) for the integral over $\partial U \cap \mathcal{D}$, we find that

$$\frac{\mathrm{d}}{\mathrm{d}t}\int_U \rho_t\,\mathrm{d}^3r = -\int_{\partial U\setminus\mathcal{D}} \vec{n}'\cdot\vec{j}\restriction_{\partial\mathcal{U}}\ \mathrm{d}S' - \int_{\partial U\cap\mathcal{D}} \vec{n}\cdot\vec{j}\restriction_{\mathcal{D}}\ \theta\big(\vec{n}\cdot\vec{j}\restriction_{\mathcal{D}}\big)\mathrm{d}S. \tag{30}$$

The first summand gives the amount of probability flowing into the other spatial region $\Omega \setminus U$, while the second summand gives the amount of probability flowing into the relevant part of the detector surface, $\partial U \cap \mathcal{D}$. That the latter is negative for arbitrary choices of $U$ means that probability can only flow into the detector surface $\mathcal{D}$; it can never flow out.

In summary, it was shown that the modified continuity Equation (23) in conjunction with the source/sink function $R$, as given via Equation (28), assures that the detector is absorbing. Conversely, if this is assumed, then probability conservation yields that the rate of change for the surface probability density $\sigma$ is given via Equation (26).

Since this was achieved without the introduction of any restrictive boundary condition, Equations (23), (28), and (26) may be understood as a general detector model. To set some terminology, we say that those equations model an *ideal detector (with surface probability density* σ*)*.

## 4. Dynamics in the presence of an ideal detector

In Section 3 above, we have seen that one can model a perfectly absorbing ideal detector by adding an appropriate sink term to the continuity equation. However, the Schrödinger equation implies the continuity Equation (10) (cf. Section 2). The Schrödinger equation is therefore incompatible with the derived, modified continuity Equation (23) for $R \neq 0$.

In this section we argue that the modified continuity Equation (23) may still be used to set up physically sensible dynamical equations. This is achieved by taking a broader view on quantum dynamics, which is based on a system of partial differential equations discovered by Madelung in the 1920s [20–22,27,34]. While the resulting, adapted system of dynamical equations is incompatible with the quantum-mechanical formalism, it may be regarded as a typical model within *geometric quantum theory*. Geometric quantum theory is a novel adaption of quantum mechanics, which makes the latter consistent with mathematical probability theory (cf. [28,34]). After motivating the dynamical equations of interest here, we provide a brief review of the theory and show how the results of Section 3 are naturally embedded within this novel formalism.

In 1926 Madelung derived a reformulation of Schrödinger's equation [20–22]. His system of PDEs is closely reminiscent of dynamical equations commonly encountered in Newtonian continuum mechanics. A sizable amount of literature on the Madelung equations may be found under the somewhat deceiving heading of 'quantum hydrodynamics' (see e.g. [52–63]). Upon defining $\rho$, $\vec{v}$ and $\vec{u}$ as in Equations (6), (12), and (14) above and using a formulation due to Nelson [27], the Madelung equations are the system of PDEs consisting of the Madelung force equation (for some given potential $V$)

$$m\left(\frac{\partial \vec{v}}{\partial t} + (\vec{v} \cdot \nabla)\vec{v}\right) = -\nabla V + m(\vec{u} \cdot \nabla)\vec{u} + \frac{\hbar}{2}\Delta \vec{u}, \tag{31}$$

the continuity Equation (10), as well as the irrotationality condition

$$\nabla \times \vec{v} = 0. \tag{32}$$

The Madelung equations are a nonlinear system of constrained evolution equations in terms of the (time-dependent) probability density $\rho$ and the (time-dependent) vector field $\vec{v}$. It is a subject of contemporary mathematical research, whether there exists a distributional formulation for suitably chosen initial data such that the corresponding initial value problem is well-posed. In particular, the status of Equation (32) has been a source of much debate. The interested reader is referred to [29,52,53,55,64–66].

The Madelung equations are of interest to the discussion here, since they decouple what is arguably the dynamical part of the Schrödinger equation, Equation (31), from the part ensuring probability conservation, Equation (10).

The central idea for defining the dynamics for a single (not necessarily free) body in the presence of an ideal detector is therefore to replace Equation (10) by Equation (23) (together with Equation (28)), while keeping Equations (31) and (32). For $V = 0$, the full system of equations is then indeed given by Equation (2a). In either case, $\sigma$ may be obtained from Equation (27).

One may ask, whether this new system of PDEs gives rise to a modified Schrödinger equation. For non-distributional source terms and under stronger conditions of regularity, a corresponding modified Schrödinger equation was indeed given in Proposition 6.1 of [34].

By carrying this equation over in a purely formal manner, we obtain

$$\mathrm{i}\hbar\frac{\partial}{\partial t}\Psi = -\frac{\hbar^2}{2m}\Delta\Psi + V\Psi + \mathrm{i}\frac{\hbar^2}{2m}\vec{n}\cdot \mathrm{Im}\left(\frac{\nabla\Psi}{\Psi}\Big|_{\mathcal{D}}\right)\theta\big(-\vec{n}\cdot \mathrm{Im}\big(\Psi^*\nabla\Psi\big)\restriction_{\mathcal{D}}\big)\delta_{\mathcal{D}}\Psi\restriction_{\mathcal{D}}. \quad (33)$$

Whether and how Equation (33) may be recast into a mathematically meaningful expression is an open question. The interested reader may find an arguably natural approach to this question in the appendix.

Quantum mechanics requires that the infinitesimal generator of time evolution, the Hamiltonian, is a linear and even (unbounded) self-adjoint operator on the respective $L^2$-space (cf. Sec. VIII.4 in [67] and Section 1.2.1 in [68]). Equation (33) is nonlinear in $\Psi$ and the corresponding Hamiltonian fails to be even symmetric. Therefore, at least within the theory of quantum mechanics, this detector model leads to unacceptable dynamics.

Still, Equations (23), (28), (31), (32), and (26) (respective Equation (2a) for $V = 0$) in principle define a sensible system of PDEs for the quantities $\rho$, $\vec{v}$, and $\sigma$. We shall call this system of equations the *(1-body) Madelung equations in the presence of an ideal detector*. Of course, mathematically one still requires an appropriate weak formulation and a proof of well-posedness for suitably chosen initial data [29], yet such questions are beyond the scope of this work. Rather, we shall ask whether the system is acceptable from a physical point of view, in spite of the overt incompatibility with the theory of quantum mechanics.

It was already argued that this dynamical model is physically natural. Yet if it is not embedded within a physically justified theoretical framework, it may still seem ad hoc. This theoretical framework exists and is known as *geometric quantum theory*. Indeed, the solution of the time of arrival problem presented in this work ought to be considered an application of geometric quantum theory – one in which the conceptual differences to the theory of quantum mechanics exhibit themselves on a dynamical level. In Section 7 of [28], the existence of such models was already anticipated.

Geometric quantum theory is a research program in both non-relativistic and relativistic quantum theory, which is based on two core hypotheses:

(I) Fundamental equations of quantum theory admit a Madelung formulation (i.e. a formulation akin to the Madelung equations).
(II) Quantum phenomena are described via mathematical probability theory.

Originally, geometric quantum theory was born out of the observation that the Madelung equations suggest a physically natural solution to the problem of (first) quantisation under holonomic constraints (cf. Section 1.1 and Remark 3.5 in [34]). Instead of providing yet another ad hoc quantisation scheme, however, geometric quantum theory traces the common quantum-mechanical operators back to physically natural random variables. That is, at each time ‘observables’ are modelled by real scalar or real vector-valued functions on the probability space. Those random variables are motivated from the close analogy of the Madelung equations to the theory of diffusion (cf. Section 4 in [34] and Section 5 in [28]). For the many-body Schrödinger theory with time-independent scalar potential, it was found that central quantum-mechanical predictions are reproduced in this manner. Moreover, for this class of models a mathematically rigorous formulation has been attained and the ability of the theory to make novel predictions in certain instances has been ascertained [28]. Geometric quantum theory provides a natural approach to the classical limit, which is asymptotically given by the case that $\hbar$ is small in comparison to the characteristic length and time scales in the particular model (cf. [34,55] and Rem. 3.1 in [29]). Roughly, this is the limit in which the latter two summands on the right hand side of Equation (31) vanish. Thus, in geometric quantum theory there is no 'Heisenberg cut' and in passing to the classical limit no need arises to change the underlying probability theory. Furthermore, an approach to (general-)relativistic geometric quantum theory has already been developed in [69].

Geometric quantum theory is connected to de Broglie-Bohm theory [70–74] and Nelson's theory of stochastic mechanics [27,75–78] through the incorporation of the Madelung equations into the basic formulation of the theory. Still, the latter two theories are mostly depicted as reinterpretations or extensions of quantum mechanics, while geometric quantum theory is a distinct theory. Of course, if the current velocity vector field is sufficiently regular, the Picard-Lindelöf theorem makes it possible to consider 'Bohm trajectories' within geometric quantum theory. Furthermore, in Section 2.2 of [35] it was argued that, for reasons of internal consistency, stochastic mechanics requires mathematical probability theory to hold. If this view is taken, geometric quantum theory constitutes a generalisation of stochastic mechanics, which does not rely on any particular model of particle trajectories (see also [79,80]).

In practical applications of geometric quantum theory, one carries over the respective Schrödinger equation from quantum mechanics and one then considers its Madelung formulation. By Postulate (I), geometric quantum theory requires that such a formulation exists. For the $N$-body Schrödinger equation [35,74], the Pauli equation [54,61] and the Dirac equation [81–83] such formulations have indeed been discovered. In the next step, one replaces the quantum-mechanical operators by respective physical random variables, as motivated from the Madelung formulation. In accordance with Postulate (II), this allows one to employ mathematical probability theory for making specific predictions in relation to the probability distributions of those random variables (cf. Section 7 in [28]).

We shall discuss the implications of geometric quantum theory for the class of models considered in this work. Following [28], a physical model in geometric quantum theory consists of three basic ingredients:

(1) A time-dependent probability space, which gives the probability that the bodies are in particular configurations (Born rule for position).
(2) A collection of time-dependent physical random variables, such as position, energy, etc.
(3) A time evolution law for those quantities.

For the problem at hand, the probability space has already been described in Section 3. Through the use of differential equations and initial conditions for $\rho$ and $\sigma$, we were able to guarantee that condition (16) is met for all times.

As random variables, we have introduced the probability current velocity $\vec{v}$ and the stochastic velocity $\vec{u}$ (via $\rho$ in Equation (1)). The position random variable $\vec{r}$, the probability current momentum $\vec{p} = m\vec{v}$ as well as the angular momentum of the probability current $\vec{r} \times \vec{p}$ (around the origin) were implicit in the description. The energy random variable can be directly obtained from the Madelung force equation:

$$E = \frac{m}{2}\vec{v}^2 + V - \frac{m}{2}\vec{u}^2 - \frac{\hbar}{2}\nabla \cdot \vec{u}. \tag{34}$$

Note that the latter two summands are responsible for quantum tunnelling.

In the case of interest, $V = 0$ and thus the time evolution of $\rho$, $\sigma$ and $\vec{v}$ is given via the free Madelung equations in the presence of an ideal detector, Equation (2a). From those three quantities all other physically relevant random variables can be computed.

We conclude this section with the note that, if $\mathbb{P}_t(\mathcal{D}) \neq 0$ for some $t$, we may compute the conditional probability measure given $\mathcal{D}$ (at time $t$). It is denoted by $\mathbb{P}_t(\,.\,|\mathcal{D})$ and for any $U \in \mathcal{B}^*(\Omega)$ it is given via

$$\mathbb{P}_t(U|\mathcal{D}) = \frac{\mathbb{P}_t(U \cap \mathcal{D})}{\mathbb{P}_t(\mathcal{D})} = \frac{1}{\mathbb{P}_t(\mathcal{D})}\int_{U\cap\mathcal{D}} \sigma_t \,\mathrm{dS} \tag{35}$$

(cf. Definition 8.2 in [17]). $\mathbb{P}_t(\,.\,|\mathcal{D})$ is of interest, because it allows us to compute probabilities, probability distributions and expectation values of random variables under the condition that the body is on the detector screen. This may be considered an 'after/during measurement' condition.

For instance, the expectation of the probability current momentum at time $t$ with respect to $\mathbb{P}_t(\,.\,|\mathcal{D})$ is given via

$$\mathbb{E}_t(\vec{p}_t|\mathcal{D}) = \frac{1}{\mathbb{P}_t(\mathcal{D})}\int_{\mathcal{D}} \vec{p}_t \restriction_{\mathcal{D}} \sigma_t \,\mathrm{dS} \tag{36}$$

(cf. Definition 8.9 in [17]). It may be understood as a measure of how 'hard' the body hits the detector on average at time $t$.

## 5. Time of arrival probability

In order to solve the time of arrival problem, we now ask for the probability that the body impacts on the detector surface between $t$ and $t + \Delta t$ with $t \geq 0$ and $\Delta t > 0$. We shall assume that initially $\mathbb{P}_0(\mathcal{D}) = 0$, which is equivalent to $\sigma_0 = 0$ (Lebesgue almost everywhere).

To begin, it is imperative to know that neither quantum mechanics nor geometric quantum theory in its current state of development are able to make specific predictions for individual bodies: they are concerned with the ensemble [31–33] as a whole, not the individual samples constituting the ensemble (cf. Section 2.1 in [35] and Section 3 in [28]). They are phenomenological theories in the sense that only probabilities for the ensemble can be computed. For instance, they can predict the probability that the body is in a certain region of space at time $t$; yet they can generally not predict that the body is in a certain region of space at time $t + \Delta t$, given that *the same body* was in some other region at time $t$. In other words, there is generally no 'resolution' on the level of the individual sample, as so called 'hidden variable theories' such as de Broglie-Bohm theory [12,45] or stochastic mechanics [11] aim to achieve (see also [84] for the related concept of 'first hit time' in the theory of stochastic processes).

That the context at hand provides us with a different situation is due to the fact that the detector model was constructed in such a manner, that after impact the individual body stays at its impact location.

Thus, we can infer that for an ideal detector the probability that the body hits the detector surface up until (and including) any time $t' \geq 0$ is given by the quantity

$$\mathbb{T}\big([0, t']\big) = \mathbb{P}_{t'}(\mathcal{D}). \tag{37}$$

Accordingly, Equation (37) does not hold, when the detector is not absorbing, i.e. when there is so called 'quantum reentry' [11,48,49]: in general, $\mathbb{P}_{t'}(\mathcal{D})$ gives the probability that at time $t'$ a given body in the ensemble is on the detector surface. Hence, if that body can leave its location of impact, $\mathbb{P}_{t'}(\mathcal{D})$ does not contain any information about prior impacts. It is only through the ideal detector model that we can use $\mathbb{P}$ to determine the impact time distribution. In this instance, this distribution is the same as the distribution for the 'time of flight', defined as the time the body stays in the interior $\Omega$ after emission at $t = 0$.

To determine the arrival probability for an interval, we formally require $\mathbb{T}$ to be a probability measure. The function $t' \mapsto \mathbb{T}([0, t'])$ on $[0, \infty)$, as given by Equation (37), may then be formally viewed as cumulative probability distribution function. Hence, we find

$$\mathbb{T}((t, t + \Delta t]) = \mathbb{T}([0, t + \Delta t]) - \mathbb{T}([0, t]) \tag{38}$$

$$= \mathbb{P}_{t+\Delta t}(\mathcal{D}) - \mathbb{P}_t(\mathcal{D}). \tag{39}$$

While cumulative distribution functions may be discontinuous in general, the formula for the rate $\partial\sigma/\partial t$, Equation (26), suggests that $t' \mapsto \mathbb{T}([0, t'])$ is continuous. Therefore, $\mathbb{T}(\{t'\}) = 0$ for any $t' \geq 0$ and thus

$$\mathbb{T}([t, t + \Delta t]) = \mathbb{T}((t, t + \Delta t]) = \mathbb{P}_{t+\Delta t}(\mathcal{D}) - \mathbb{P}_t(\mathcal{D}). \tag{40}$$

Equation (40) provides the answer to the stated problem. An analogous expression was given by Daumer et al. [45], though using the different rate (21).

We shall also construct the respective probability space $(\Lambda, \mathcal{A}', \mathbb{T})$. First, we recall that the function $t \mapsto \mathbb{P}_t(\mathcal{D})$ is monotonously increasing and bounded from above by 1. The limit

$$p_\infty = \lim_{t\to\infty} \mathbb{P}_t(\mathcal{D}) \tag{41}$$

therefore exists, yet it need not equal 1. By definition, $p_\infty$ is the probability that the body impacts on the detector surface at an any time:

$$\mathbb{T}([0, \infty)) = p_\infty. \tag{42}$$

Conversely, the quantity $1 - p_\infty$ gives the probability that the body never reaches the detector surface. Unless $p_\infty = 1$, one needs to account for this particular event in defining $\mathbb{T}$ (cf. [1,5]). We shall choose the convention to symbolically represent the event by the singleton $\{\emptyset\}$. We therefore obtain the following

$$\Lambda = [0, \infty) \cup \{\emptyset\} \tag{43}$$

with

$$\mathbb{T}(\{\emptyset\}) = 1 - p_\infty \quad \text{and} \quad \mathbb{T}(\Lambda) = 1. \tag{44}$$

Accordingly, the $\sigma$-algebra $\mathcal{A}'$ is the one generated from the union of the Lebesgue sets $\mathcal{B}^*([0,\infty))$ and the set $\{\emptyset\}$ (cf. Theorem 1.16 in [17]). We shall write this as follows:

$$\mathcal{A}' = \Sigma\big(\mathcal{B}^*([0,\infty)) \cup \{\emptyset\}\big). \tag{45}$$

Equation (40) then suggests the following general formula for the probability measure $\mathbb{T}$ for an arbitrary $J \in \mathcal{A}'$:

$$\mathbb{T}(J) = \int_{J\setminus\{\emptyset\}} \mathrm{d}t \int_{\mathcal{D}} \mathrm{dS}\, \frac{\partial\sigma}{\partial t} + \mathbb{T}(J \cap \{\emptyset\}). \tag{46}$$

Formula (46) in conjunction with Equation (26) (along with the respective probability space) is indeed what is suggested here as a general solution to the time of arrival problem for a single body and an ideal detector. $\mathbb{T}$ is the probability distribution for the impact time of the body, which also accounts for the event of no impact. For $J \in \mathcal{B}^*([0,\infty))$ the quantity $\mathbb{T}(J)$ is the probability that the body impacts on the detector at any time $t$ with $t \in J$. For $J = \{\emptyset\}$ the quantity $\mathbb{T}(J) = \mathbb{T}(\{\emptyset\})$ is the probability that the body never impacts on the detector.

To close this section, we shall introduce a closely related probability space, which allows one to compute the probability that the body impinges on the surface in a given interval of time and a given region on the detector surface. Determining this probability is called the *screen problem* [23,30].

The respective probability measure $\mathbb{D}$ can be defined on the measurable space

$$\Big(([0,\infty) \times \mathcal{D}) \cup \{\emptyset\}, \Sigma\big(\mathcal{B}^*([0,\infty) \times \mathcal{D}) \cup \{\emptyset\}\big)\Big). \tag{47}$$

For the event of no impact we again have

$$\mathbb{D}(\{\emptyset\}) = 1 - p_\infty. \tag{48}$$

For an event $U \in \mathcal{B}^*(\mathcal{D} \times [0,\infty))$ the measure is given via

$$\mathbb{D}(U) = \int_U \mathrm{d}t\mathrm{dS}\, \frac{\partial\sigma}{\partial t}. \tag{49}$$

In analogy to Equation (46), we may therefore write down the general formula

$$\mathbb{D}(J) = \int_{J\setminus\{\emptyset\}} \mathrm{d}t\mathrm{dS}\, \frac{\partial\sigma}{\partial t} + \mathbb{D}(J \cap \{\emptyset\}) \tag{50}$$

for any $J \in \Sigma(\mathcal{B}^*([0,\infty) \times \mathcal{D}) \cup \{\emptyset\})$.

We call $\mathbb{D}$ the *(ideal) detector distribution.* $\mathbb{D}$ is the joint probability distribution for the impact location and the impact time of the body, which also accounts for the event of no impact. For $J$ as above with $J \cap \{\emptyset\} = \emptyset$, the quantity $\mathbb{D}(J)$ is the probability that the body impacts on the detector at any time $t$ and location $\vec{r}$ with $(t,\vec{r}) \in J$. For $J = \{\emptyset\}$ the quantity $\mathbb{D}(J) = \mathbb{D}(\{\emptyset\}) = \mathbb{T}(\{\emptyset\})$ is the probability that the body never impacts on the detector. As noted by Cavendish and Das [23], $\mathbb{T}$ may be viewed as a marginal distribution of $\mathbb{D}$.

This solution of the screen problem via Equation (50) is, of course, provided in the context of the dynamics determined by Equation (2a). Otherwise, the solution is not wholly original, for, again, Daumer et al. [45] have implicitly suggested Equation (49) to hold and a more explicit formula was given by Tumulka in [13] – though both works used Equation (21) instead of Equation (26) for the rate.

## 6. Conclusion

In the context of the time of arrival problem, we have addressed the question of measurement by explicitly including the ideal detector model in the equations for the dynamical evolution of the system. In the context of the larger, foundational debate on measurement within quantum theory, it shall be remarked that measurement always requires a probe, such as a detector, another particle or radiation. That contemporary theory, at least in part, replaces this physical process by an abstract projection instead of modelling it directly (cf. [31,85] and Remark 4.55(a) in [68]) is arguably one of the reasons why a resolution of the time of arrival problem has proven so difficult: therein lies the motivation for the numerous attempts to construct a 'time operator'. Arguably, such approaches did not succeed, because they pursued abstraction on the grounds of weak physical arguments over concrete physical models [1,30]. In this sense, the aforementioned, paradoxical status of the time of arrival problem in contemporary quantum theory points at the limitations of the projection postulate in our contemporary theory of measurement. Addressing these limitations will require, on the one hand, a justification for the empirical success of the projection postulate in certain instances, and, on the other hand, a theoretical framework, which does not rely on it for describing the physical process of measurement.

This work constitutes a step in this direction. Here it was possible to forego the use of the projection postulate by modelling the measurement outcomes directly via the probability surface density $\sigma$ Section 3. The respective detector model, however, necessitates a broader view on quantum dynamics than what the theory of quantum mechanics can supply. Far from being ad hoc, it was shown in Section 4 how this broader view on quantum dynamics is embedded within the theoretical framework of geometric quantum theory [28,34,35]. Geometric quantum theory is a novel adaption of quantum mechanics, which makes the latter consistent with mathematical probability theory (cf. [28,34]). In the theory, predictions of measurement outcomes are obtained by including the interaction of the probe with the system of interest in the dynamical model, rather than relying on the Dirac-von Neumann axioms. Since geometric quantum theory generally uses the same evolution equations for modelling this process as quantum mechanics, differences in prediction tend to be quantitative in nature, seldom qualitative.

At the end of Section 4, it was also shown how geometric quantum theory is able to make statements about the system 'after measurement': for this purpose, mathematical probability theory supplies the tools of conditional probability measures and conditional expectation values. It is to be expected that a general treatment, which addresses the aforementioned limitations of the projection postulate, will also make use of these tools.

In this respect, it is also worthwhile to recall the limitations of both quantum mechanics and geometric quantum theory in its current form with regards to making predictions for individual samples in the ensemble Section 5: only in isolated instances, it is possible to make such predictions. In this instance, it was the ideal detector model that ultimately enabled the prediction of the time of impact distribution. For the related tunnelling time problem [86–89], however, nothing can be said about the position of the individual body shortly after it reaches or has traversed the barrier. That is, without any further physical assumptions, this problem is beyond the limitations of contemporary theory.

We shall also address further developments of the model presented in this work. Among those, finding an appropriate distributional formulation of the system of Equation (2a) appears to be most pressing. A contribution towards the resolution of this question for the original Madelung equations may be found in [29], though even for that system the question remains unresolved and a proof of well-posedness for an appropriate formulation with suitably chosen initial data $(\rho_0, \vec{v}_0)$ currently seems to be beyond reach. Well-posedness results for non-linear systems of equations in Newtonian continuum mechanics are, however, known to be difficult to obtain, so that a more pragmatic approach seems appropriate. Numerical solutions to a given distributional formulation of Equation (2a) may be possible without a rigorous result of well-posedness. In lower-dimensional analogue models, closed-form solutions may be within reach. Furthermore and as done in the appendix, one may consider a suitable weak formulation of the modified Schrödinger equation (33) instead. Such pragmatic approaches may also open up the path to empirical tests.

An obvious further refinement of the presented model is to include particle spin and its interaction with electromagnetic radiation [54,61,89–94]. Works that aim to solve the time of arrival problem for multiple bodies already exist [42,95–97], so it is natural to ask how to generalise Equation (2a) to the many-body case. The latter is also of potential, independent concern, since the ability to enclose a finite domain with a perfectly absorbing boundary may have applications to numerical quantum chemistry (see e.g. [98] on the so called 'curse of dimensionality' in this context). Another possible extension of the model is the inclusion of interactions of the body with the detector surface: for instance, one may charge the detector and consider electromagnetic interactions or even include stochastic interactions with the detector surface, thus creating a potential link to the theory of open quantum systems [99,100]. Of course, it is also natural to consider generalisations to the relativistic domain [69,97,101,102].

On a final note, one may extend the model to contexts, which are of central importance to the foundations of quantum theory. For instance, one may consider the ideal detector model in the context of Bell test experiments [93,103,104]. It may also be used to obtain a probabilistic 'near field' model of the double slit experiment [39,40,105–107] by modelling the screen with the slits via a homogeneous Dirichlet boundary condition. Such a model may indeed provide an illustrative example for how geometric quantum theory, which is based on mathematical probability theory and a point mass model of particles, is able to predict the correct wave-like statistics.

## Acknowledgments

The author would like to thank Valter Moretti for drawing his attention to the problem as well as Pascal Naidon, Siddhant Das, Ángel Sanz, Sophya Garashchuk, Vitaly Rassolov, Lev Vaidman, and Alisson Tezzin for helpful discussion.

## Disclosure statement

No potential conflict of interest was reported by the author(s).

## ORCID

*Maik Reddiger* http://orcid.org/0000-0002-0485-5044

## References

[1] G.R. Allcock, *The time of arrival in quantum mechanics I. Formal considerations*, Ann. Phys. 53 (1969), pp. 253–285. https://doi.org/10.1016/0003-4916(69)90251-6.
[2] W. Pauli, *Die allgemeinen prinzipien der wellenmechanik*, in *Prinzipien der Quantentheorie I*, S. Flügge, ed., Springer, Vol. 5, 1958, pp. 1–160.
[3] W. Pauli, *Die allgemeinen Prinzipien der Wellenmechanik: Neu heraus gegeben und mit historischen Anmerkungen versehen von Norbert Straumann*, Springer, Berlin, 1990.
[4] R. Brunetti and K. Fredenhagen, *Time of occurrence observable in quantum mechanics*, Phys. Rev. A 66 (2002), pp. 044101. https://doi.org/10.1103/PhysRevA.66.044101.
[5] S. Das and W. Struyve, *Questioning the adequacy of certain quantum arrival-time distributions*, Phys. Rev. A 104 (2021), pp. 042214. https://doi.org/10.1103/PhysRevA.104.042214.
[6] V. Delgado and J.G. Muga, *Arrival time in quantum mechanics*, Phys. Rev. A 56 (1997), pp. 3425–3435. https://doi.org/10.1103/PhysRevA.56.3425.
[7] S. Goldstein, R. Tumulka, and N. Zanghì, Arrival times versus detection times. Found. Phys. 54 (2024), article number 63. https://doi.org/10.1007/s10701-024-00798-y.
[8] N. Grot, C. Rovelli, and R.S. Tate, *Time of arrival in quantum mechanics*, Phys. Rev. A 54 (1996), pp. 4676–4690. https://doi.org/10.1103/PhysRevA.54.4676.
[9] T. Jurić and H. Nikolić, *Arrival time from Hamiltonian with non-Hermitian boundary term*, Universe 10 (2024), article number 35. https://doi.org/10.3390/universe10010035.
[10] J. Kijowski, *On the time operator in quantum mechanics and the heisenberg uncertainty relation for energy and time*, Rep. Math. Phys. 6 (1974), pp. 361–386. https://doi.org/10.1016/S0034-4877(74)80004-2.
[11] P. Naidon, *Inequivalence of stochastic and Bohmian arrival times in time-of-flight experiments*, Phys. Rev. A 109 (2024), pp. 063312. https://doi.org/10.1103/PhysRevA.109.063312.
[12] M.V. Scherer, A.D. Ribeiro, and R.M. Angelo, *Testing trajectory-based determinism via probability distributions*. Chin. J. Phys 97 (2025), pp. 1199–1212. https://doi.org/10.1016/j.cjph.2025.08.020.
[13] R. Tumulka, *Distribution of the time at which an ideal detector clicks*, Ann. Phys. 442 (2022), pp. 168910. https://doi.org/10.1016/j.aop.2022.168910.
[14] R. Werner, *Arrival time observables in quantum mechanics*, Ann. Inst. Henri Poincaré 47 (1987), pp. 429–449.
[15] J.G. Muga and C.R. Leavens, *Arrival time in quantum mechanics*, Phys. Rep. 338 (2000), pp. 353–438. https://doi.org/10.1016/S0370-1573(00)00047-8.
[16] S. Das, *Arrival-time distributions and spin in quantum mechanics*, PhD thesis, Ludwig-Maximilians-University of Munich, Munich, 2023.
[17] A. Klenke, Probability Theory: A Comprehensive Course, 3rd ed., Springer, Berlin, 2020. https://doi.org/10.1007/978-3-030-56402-5.
[18] A. Kolmogoroff, Grundbegriffe der Wahrscheinlichkeitsrechnung, Springer, Berlin, 1933. https://doi.org/10.1007/978-3-642-49888-6.
[19] A. Kolmogorov, Foundations of the Theory of Probability, 2nd ed., Chelsea Publishing, New York, 1956. Trans. by N. Morrison.
[20] E. Madelung, *Eine anschauliche deutung der gleichung von Schrödinger*, Naturwissenschaften 14 (1926), pp. 1004–1004. https://doi.org/10.1007/BF01504657.

[21] E. Madelung, *Quantentheorie in hydrodynamischer form*, Z. Physik 40 (1927), pp. 322–326. https://doi.org/10.1007/BF01400372.
[22] E. Madelung, *Quantum theory in hydrodynamical form*. Trans. by D. Delphenich, 2015. Available at http://www.neo-classical-physics.info/uploads/3/4/3/6/34363841/madelung_-_hydrodynamical_interp..pdf.
[23] W. Cavendish and S. Das, *Absorbing detectors meet scattering theory*, Phys. Rev. A 112 (2025), pp. 062217. https://doi.org/10.48550/arXiv.2509.07518.
[24] M. Reed and B. Simon, Methods of Modern Mathematical Physics: Scattering Theory, Vol. III, Academic Press, New York, 1979.
[25] J.R. Taylor, Scattering Theory: The Quantum Theory on Nonrelativistic Collisions, Wiley, New York, 1972.
[26] E.A. Carlen, *Conservative diffusions*, Commun. Math. Phys. 94 (1984), pp. 293–315. https://doi.org/10.1007/BF01224827.
[27] E. Nelson, *Derivation of the Schrödinger equation from Newtonian mechanics*, Phys. Rev. 150 (1966), pp. 1079–1085. https://doi.org/10.1103/PhysRev.150.1079.
[28] M. Reddiger, *On the applicability of Kolmogorov's theory of probability to the description of quantum phenomena. Part I: Foundations*, Quantum Stud.: Math. Found. 13 (2026), article number 1. https://doi.org/10.1007/s40509-025-00375-6.
[29] M. Reddiger and B. Poirier, *Towards a mathematical theory of the Madelung equations: Takabayasi's quantization condition, quantum quasi-irrotationality, weak formulations, and the Wallstrom phenomenon*, J. Phys. A: Math. Theor. 56 (2023), pp. 193001. https://doi.org/10.1088/1751-8121/acc7db.
[30] B. Mielnik, *The screen problem*, Found. Phys. 24 (1994), pp. 1113–1129. https://doi.org/10.1007/BF02057859.
[31] L.E. Ballentine, *The statistical interpretation of quantum mechanics*, Rev. Mod. Phys. 42 (1970), pp. 358–381. https://doi.org/10.1103/RevModPhys.42.358.
[32] A. Pechenkin, *The statistical (ensemble) interpretation of quantum mechanics*, in *The Oxford Handbook of the History of Quantum Interpretations*, O. Freire Junior, G. Bacciagaluppi, O. Darrigol, T. Hartz, C. Joas, A. Kojevnikov and O. Pessoa Junior, eds., Oxford University Press, Oxford, 2022, pp. 1247–1264.
[33] R. von Mises, Probability, Statistics, and Truth, 2nd ed., Dover Publications, New York, 1957.
[34] M. Reddiger, *The Madelung picture as a foundation of geometric quantum theory*, Found. Phys. 47 (2017), pp. 1317–1367. https://doi.org/10.1007/s10701-017-0112-5.
[35] M. Reddiger, *Towards a probabilistic foundation for non-relativistic and relativistic quantum theory*, PhD thesis, Lubbock: Texas Tech University, 2022. url: https://hdl.handle.net/2346/91876
[36] T. Fevens and H. Jiang, *Absorbing boundary conditions for the Schrödinger equation*, SIAM J. Sci. Comput. 21 (1999), pp. 255–282. https://doi.org/10.1137/S1064827594277053.
[37] P.C. Bressloff, *Diffusion-mediated absorption by partially-reactive targets: Brownian functionals and generalized propagators*, J. Phys. A: Math. Theor. 55 (2022), pp. 205001. https://doi.org/10.1088/1751-8121/ac5e75.
[38] L. Frolov, *Detecting screens modeled by Schrödinger operators that generate C0 contraction semigroups*, Rev. Math. Phys. 11 (2025), article number 2550036. https://doi.org/10.1142/S0129055X25500369.
[39] A. Ayatollah Rafsanjani, M. Kazemi, A. Bahrampour, and M. Golshani, *Can the double-slit experiment distinguish between quantum interpretations?* Commun. Phys. 6 (2023), article number 195. https://doi.org/10.1038/s42005-023-01315-9.
[40] A. Ayatollah Rafsanjani, M. Kazemi, V. Hosseinzadeh, and M. Golshani, *Non-local temporal interference*, Sci. Rep. 14 (2024), article number 3615. https://doi.org/10.1038/s41598-024-54018-8.
[41] L. Frolov, S. Teufel, and R. Tumulka, *Existence of Schrödinger evolution with absorbing boundary condition*, Math. Phys. Anal. Geom. 28 (2025), pp. 27. https://doi.org/10.1007/s11040-025-09521-3.

[42] R. Tumulka, *Detection-time distribution for several quantum particles*, Phys. Rev. A 106 (2022), pp. 042220. https://doi.org/10.1103/PhysRevA.106.042220.
[43] R. Tumulka, *Absorbing boundary condition as limiting case of imaginary potentials*, Commun. Theor. Phys. 75 (2022), pp. 015103. https://doi.org/10.1088/1572-9494/ac9bea.
[44] R. Tumulka, *On a derivation of the absorbing boundary rule*, Phys. Lett. A 494 (2024), pp. 129286. https://doi.org/10.1016/j.physleta.2023.129286.
[45] M. Daumer, D. Dürr, S. Goldstein, and N. Zanghì, *On the quantum probability flux through surfaces*, J. Stat. Phys. 88 (1997), pp. 967–977. https://doi.org/10.1023/B:JOSS.0000015181.86864.fb.
[46] N. Vona and D. Dürr, *The role of the probability current for time measurements*, in *The Message of Quantum Science: Attempts Towards a Synthesis*, P. Blanchard and J. Fröhlich, eds., Springer, Heidelberg, 2015, pp. 95–112.
[47] N. Vona, G. Hinrichs, and D. Dürr, *What does one measure when one measures the arrival time of a quantum particle?* Phys. Rev. Lett. 111 (2013), pp. 220404. https://doi.org/10.1103/PhysRevLett.111.220404.
[48] A. Goussev, *Equivalence between quantum backflow and classically forbidden probability flow in a diffraction-in-time problem*, Phys. Rev. A 99 (2019), pp. 043626. https://doi.org/10.1103/PhysRevA.99.043626.
[49] A. Goussev, *Probability backflow for correlated quantum states*, Phys. Rev. Res. 2 (2020), pp. 033206. https://doi.org/10.1103/PhysRevResearch.2.033206.
[50] G.R. Allcock, *The time of arrival in quantum mechanics III. The measurement ensemble*, Ann. Phys. 53 (1969), pp. 311–348. https://doi.org/10.1016/0003-4916(69)90253-X.
[51] M.J. Schervish, Theory of Statistics, 2nd ed., Springer, New York, 1997.
[52] P. Antonelli, L.E. Hientzsch, and P. Marcati, *On some results for quantum hydrodynamical models (Mathematical analysis in fluid and gas dynamics)*, in 数理解析研究所講究録 2070, 2018, pp. 107–129. Available at http://hdl.handle.net/2433/241991.
[53] P. Antonelli and P. Marcati, *On the finite energy weak solutions to a system in quantum fluid dynamics*, Commun. Math. Phys. 287 (2009), pp. 657–686. https://doi.org/10.1007/s00220-008-0632-0.
[54] I. Bialynicki-Birula, M. Cieplak, and J. Kaminski, Theory of Quanta, Oxford University Press, New York, 1992. Trans. by A. M. Furdyna.
[55] I. Gasser and P.A. Markowich, *Quantum hydrodynamics, Wigner transforms and the classical limit*, Asymptot. Anal. 14 (1997), pp. 97–116. https://doi.org/10.3233/ASY-1997-14201.
[56] P. Holland, *Symmetries and conservation laws in the Lagrangian picture of quantum hydrodynamics*, in *Concepts and Methods in Modern Theoretical Chemistry: Statistical Mechanics*, S. Ghosh and P. Chattaraj, eds., Taylor & Francis, Boca Raton, 2012.
[57] L. Jánossy, *Zum hydrodynamischen modell der quantenmechanik*, Z. Physik 169 (1962), pp. 79–89. https://doi.org/10.1007/BF01378286.
[58] L. Jánossy, *The hydrodynamical model of wave mechanics: The many-body problem*, Acta Physica 27 (1969), pp. 35–46. https://doi.org/10.1007/BF03156734.
[59] L. Jánossy and M. Ziegler, *The hydrodynamical model of wave mechanics I: The motion of a single particle in a potential field*, Acta Phys. Hung. 16 (1963), pp. 37–48. https://doi.org/10.1007/BF03157004.
[60] L. Jánossy and M. Ziegler-Náray, *The hydrodynamical model of wave mechanics II: The motion of a single particle in an external electromagnetic field*, Acta Phys. Hung. 16 (1964), pp. 345–353. https://doi.org/10.1007/BF03157974.
[61] L. Jánossy and M. Ziegler-Náray, *The hydrodynamical model of wave mechanics III: Electron spin*, Acta Phys. Hung. 20 (1966), pp. 233–251. https://doi.org/10.1007/BF03158167.
[62] A. Jángel, M.C. Mariani, and D. Rial, *Local existence of solutions to the transient quantum hydrodynamic equations*, Math. Mod. Meth. Appl. S 12 (2002), pp. 485–495. https://doi.org/10.1142/S0218202502001751.
[63] L.M. Morato, *Formation of singularities in madelung fluid: A nonconventional application of itô calculus to foundations of quantum mechanics*, in *Stochastic Analysis and Applications: The Abel Symposium 2005*, Springer, 2007, pp. 527–540.

[64] T.C. Wallstrom, *On the derivation of the Schrödinger equation from stochastic mechanics*, Found. Phys. Lett. 2 (1989), pp. 113–126. https://doi.org/10.1007/BF00696108.

[65] T.C. Wallstrom, *On the initial-value problem for the Madelung hydrodynamic equations*, Phys. Lett. A 184 (1994), pp. 229–233. https://doi.org/10.1016/0375-9601(94)90380-8.

[66] T.C. Wallstrom, *Inequivalence between the Schrödinger equation and the Madelung hydrodynamic equations*, Phys. Rev. A 49 (1994), pp. 1613–1617. https://doi.org/10.1103/PhysRevA.49.1613.

[67] M. Reed and B. Simon, Methods of Modern Mathematical Physics: Functional Analysis, Vol. I, Academic Press, New York, 1972.

[68] V. Moretti, Fundamental Mathematical Structures of Quantum Theory: Spectral Theory, Foundational Issues, Algebraic Formulation, Springer, Cham, 2019.

[69] M. Reddiger and B. Poirier, *Towards a probabilistic foundation of relativistic quantum theory: The one-body Born rule in curved spacetime*, Quantum Stud.: Math. Found. 12 (2024), article number 5. https://doi.org/10.1007/s40509-024-00349-0.

[70] D. Bohm, *A suggested interpretation of the quantum theory in terms of hidden variables. I*, Phys. Rev. 85 (1952), pp. 166–179. https://doi.org/10.1103/PhysRev.85.166.

[71] D. Bohm, *A suggested interpretation of the quantum theory in terms of hidden variables. II*, Phys. Rev. 85 (1952), pp. 180–193. https://doi.org/10.1103/PhysRev.85.180.

[72] D. Dürr, S. Goldstein, and N. Zanghì, *On a realistic theory for quantum physics*, in *Stochastic Process, Physics and Geometry*, S. Albeverio, G. Casati, U. Cattaneo, D. Merlini, and R. Moresi, eds., World Scientific, Singapore, 1992, pp. 374–391.

[73] D. Dürr, S. Goldstein, and N. Zanghì, Quantum Physics without Quantum Philosophy, Springer, Berlin, 2013.

[74] P.R. Holland, The Quantum Theory of Motion: An Account of the de Broglie-Bohm Causal Interpretation of Quantum Mechanics, Cambridge University Press, Cambridge, 1993.

[75] G. Bacciagaluppi, *A conceptual introduction to Nelson's mechanics*, in *Endophysics, Time, Quantum and the Subjective*, World Scientific, 2005, pp. 367–388. https://doi.org/10.1142/9789812701596_0020.

[76] I. Fényes, *Eine wahrscheinlichkeitstheoretische begründung und interpretation der quantenmechanik*, Z. Physik 132 (1952), pp. 81–106. https://doi.org/10.1007/BF01338578.

[77] E. Nelson, Quantum Fluctuations, Princeton University Press, Princeton, 1985.

[78] E. Santos, *Stochastic interpretations of quantum mechanics*, in *The Oxford Handbook of the History of Quantum Interpretations*, O. Freire Junior, G. Bacciagaluppi, O. Darrigol, T. Hartz, C. Joas, A. Kojevnikov and O. Pessoa Junior, eds., Oxford University Press, Oxford, 2022, pp. 1247–1263.

[79] E.A. Carlen, *Stochastic mechanics: A look back and a look ahead*, in *Diffusion, Quantum Theory, and Radically Elementary Mathematics*, W. G. Faris, ed., Princeton University Press, Princeton, pp. 117–139. https://doi.org/10.1515/9781400865253.117.

[80] E. Nelson, *Review of stochastic mechanics*, J. Phys. Conf. Ser. 361 (2012), article number 012011. https://doi.org/10.1088/1742-6596/361/1/012011.

[81] L. Fabbri, *De Broglie–Bohm formulation of dirac fields*, Found. Phys. 52 (2022), article number 116. https://doi.org/10.1007/s10701-022-00641-2.

[82] L. Fabbri, *Dirac theory in hydrodynamic form*, Found. Phys. 53 (2023), pp. 54. https://doi.org/10.1007/s10701-023-00695-w.

[83] L. Fabbri, *Madelung structure of the Dirac equation*, J. Phys. A: Math. Theor. 58 (2025), pp. 195301. https://doi.org/10.1088/1751-8121/add2b0.

[84] R. Bass, The measurability of hitting times. Electron. Commun. Probab. 15 (2010), pp. 99–105. https://doi.org/10.1214/ECP.v15-1535.

[85] L.E. Ballentine, *Limitations of the projection postulate*, Found. Phys. 20 (1990), pp. 1329–1343. https://doi.org/10.1007/BF01883489.

[86] V. Delgado, S. Brouard, and J. Muga, *Does positive flux provide a valid definition of tunnelling times?* Solid State Commun. 94 (1995), pp. 979–982. https://doi.org/10.1016/0038-1098(95)00162-X.

[87] E.H. Hauge and J.A. Støvneng, *Tunneling times: A critical review*, Rev. Mod. Phys. 61 (1989), pp. 917–936. https://doi.org/10.1103/RevModPhys.61.917.
[88] R. Landauer and T. Martin, *Barrier interaction time in tunneling*, Rev. Mod. Phys. 66 (1994), pp. 217–228. https://doi.org/10.1103/RevModPhys.66.217.
[89] B. Poirier and R. Lombardini, *Dwell times, wavepacket dynamics, and quantum trajectories for particles with spin 1/2*, Entropy 26 (2024), article number 336. https://doi.org/10.3390/e26040336.
[90] S. Das and S. Aristarhov, *Comment on 'the spin dependence of detection times and the non-measurability of arrival times'*, 2023. https://doi.org/10.48550/arXiv.2312.01802.
[91] S. Das and D. Dürr, *Arrival time distributions of spin-1/2 particles*, Sci. Rep. 9 (2019), article number 2242. https://doi.org/10.1038/s41598-018-38261-4.
[92] S. Das, M. Nöth, and D. Dürr, *Exotic Bohmian arrival times of spin-1/2 particles: An analytical treatment*, Phys. Rev. A 99 (2019), article number 052124. https://doi.org/10.1103/PhysRevA.99.052124.
[93] A. Drezet, *Arrival time and Bohmian mechanics: It is the theory which decides what we can measure*, Symmetry 16 (2024), article number 1325. https://doi.org/10.3390/sym16101325.
[94] S. Goldstein, R. Tumulka, and N. Zanghì, *On the spin dependence of detection times and the nonmeasurability of arrival times*, Sci. Rep. 14 (2024), pp. 3775. https://doi.org/10.1038/s41598-024-53777-8.
[95] S. Selstø, *Absorption and analysis of unbound quantum particles one by one*, Phys. Rev. A 103 (2021), pp.012812. https://doi.org/10.1103/PhysRevA.103.012812.
[96] S. Selstø and S. Kvaal, *Absorbing boundary conditions for dynamical many-body quantum systems*, J. Phys. B: At. Mol. Opt. Phys. 43 (2010), pp. 065004. https://doi.org/10.1088/0953-4075/43/6/065004.
[97] A.S. Tahvildar-Zadeh and S. Zhou, *Detection time of dirac particles in one space dimension*, in *Physics and the Nature of Reality: Essays in Memory of Detlef Dürr*, A. Bassi, S. Goldstein, R. Tumulka, N. Zanghì, eds., Springer International Publishing, Cham, 2024, pp. 187–201. https://doi.org/10.1007/978-3-031-45434-9_14.
[98] V.A. Rassolov and S. Garashchuk, *Computational complexity in quantum chemistry*, Chem. Phys. Lett. 464 (2008), pp. 262–264. https://doi.org/10.1016/j.cplett.2008.09.026.
[99] H.-P. Breuer and F. Petruccione, The Theory of Open Quantum Systems, Oxford University Press, Oxford, 2002, 648 pp.
[100] A.N. Jordan and I.A. Siddiqi, Quantum Measurement: Theory and Practice, Cambridge University Press, Cambridge, 2024.
[101] S. Das, *Relativistic electron wave packets featuring persistent quantum backflow*, 2021. https://doi.org/10.48550/arXiv.2112.13180.
[102] R. Tumulka, *Detection time distribution for dirac particles*, 2016. https://doi.org/10.48550/arXiv.1601.04571.
[103] A. Aloy, T.D. Galley, C.L. Jones, S.L. Ludescher, and M.P. Mueller, *Spin-bounded correlations: Rotation boxes within and beyond quantum theory*, 2023. https://doi.org/10.48550/arXiv.2312.09278.
[104] A. Tezzin, *Violating the KCBS inequality with a toy mechanism*, Found. Sci. 30 (2025), pp. 73–87. https://doi.org/10.1007/s10699-023-09928-7.
[105] S. Das, D.-A. Deckert, L. Kellers, S. Krekels, and W. Struyve, *Double-slit experiment revisited*, Ann. Phys. 479 (2025), pp.170054. https://doi.org/10.1016/j.aop.2025.170054.
[106] G. Kälbermann, Single- and double-slit scattering of wavepackets. J. Phys. A: Math. Gen. 35 (2002), pp. 4599–4616. https://doi.org/10.1088/0305-4470/35/21/309.
[107] A.S. Sanz and S. Miret-Artés, A trajectory-based understanding of quantum interference. J. Phys. A: Math. Theor. 41 (2008), pp. 435303. https://doi.org/10.1088/1751-8113/41/43/435303.
[108] L.C. Evans, Partial Differential Equations, 2nd ed., American Mathematical Society, Providence, 2010.

## Appendix

The purpose of this appendix is to present a proposal for a weak formulation of the modified Schrödinger Equation (33). As such, it is meant to enable a rigorous study thereof and also pave the path to numerical implementations.

For an introduction to the topic of weak formulations of PDEs, the reader may consult e.g. [108], in particular Section 1.3.2. A discussion of weak formulations of the Madelung equations is given in [29].

The basic idea is to first acknowledge that Equation (33) is non-linear, hence a weak formulation of a non-linear PDE is sought. Multiplying by $\Psi^*$ is then natural for two reasons: first, division by $\Psi$ is circumvented and, second, one would also take this step in deriving the Madelung equations. One then follows the usual procedure of finding a weak formulation for a complex evolution PDE.

Accordingly, we formally multiply Equation (33) with $\Psi^*$ and rearrange:

$$\begin{aligned}0 &= \mathrm{i}\hbar\Psi^*\frac{\partial\Psi}{\partial t} + \frac{\hbar^2}{2m}\Psi^*\Delta\Psi - V\Psi^*\Psi \\ &\quad - \mathrm{i}\frac{\hbar^2}{2m}\,\vec{n}\cdot\mathrm{Im}(\Psi^*\nabla\Psi)\restriction_{\mathcal{D}}\,\theta\big(-\vec{n}\cdot\mathrm{Im}\big(\Psi^*\nabla\Psi\big)\restriction_{\mathcal{D}}\big)\delta_{\mathcal{D}}\end{aligned} \tag{A1}$$

The usual procedure is now to multiply Equation (A1) by a general, suitable ‘test function’, integrate over all variables of the test function and use integral theorems to ‘shift’ as many derivatives to the test function as possible. Since Equation (A1) is an evolution equation, we may choose one of two options: either use time-independent test functions and ask for the resulting equation to hold for almost all times or use time-dependent test functions in a ‘fully variational formulation’. We make the former choice by requiring the test function $\phi$ to be in $C_0^\infty(\Omega,\mathbb{C})$, the space of (time-independent) complex-valued, compactly supported, smooth functions on $\Omega$.

In this manner, the following weak formulation is obtained: For all $\phi \in C_0^\infty(\Omega,\mathbb{C})$

$$\begin{aligned}0 &= \int_\Omega \mathrm{d}^3 r\left(\mathrm{i}\hbar\Psi^*\frac{\partial\Psi}{\partial t} - \frac{\hbar^2}{2m}|\nabla\Psi|^2 - V|\Psi|^2 - \frac{\hbar^2}{2m}(\Psi^*\nabla\Psi)\cdot\nabla\right)\phi \\ &\quad + \int_{\mathcal{D}} \mathrm{d}S\,\frac{\hbar^2}{2m}\big(\vec{n}\cdot\mathrm{Re}(\Psi^*\nabla\Psi)\restriction_{\mathcal{D}} + \mathrm{i}\,\vec{n}\cdot\mathrm{Im}(\Psi^*\nabla\Psi)\restriction_{\mathcal{D}}\,\theta\big(+\vec{n}\cdot\mathrm{Im}(\Psi^*\nabla\Psi)\restriction_{\mathcal{D}}\big)\big)\phi\restriction_{\mathcal{D}}\end{aligned} \tag{A2}$$

holds almost everywhere on $[0,\infty)$.

It is apparent from Equation (A2) that $\Psi$ needs to be weakly differentiable at least once both in space and time. Accordingly, whenever the singular term in Equation (A1) is non-zero, it must result out of second-order distributional derivatives of $\Psi$ in space. It is also noteworthy that Equations (16) and (27) need to hold for almost all $t \in [0,\infty)$, so that the probability measure $\mathbb{P}_t$ is well-defined. Those two equations impose stronger integrability conditions on $\Psi$ and $\mathrm{Im}(\Psi^*\nabla\Psi)$ than Equation (A2) alone.

There also exists a natural approach to separating the complex Equation (A2) into two real ones. This yields a coupled system of two PDEs instead, namely an equation encoding the physical dynamics as well as a weak formulation of the continuity equation with the respective source term (2b). They are obtained by separating the real and imaginary parts of Equation (A2) for purely real (or purely imaginary) test functions. For all $\phi, \xi \in C_0^\infty(\Omega,\mathbb{R})$ we then require that the two equations

$$\begin{aligned}0 &= \int_\Omega \mathrm{d}^3 r\left(\hbar\mathrm{Im}\left(\Psi^*\frac{\partial\Psi}{\partial t}\right) + \frac{\hbar^2}{2m}|\nabla\Psi|^2 + V|\Psi|^2 + \frac{\hbar^2}{2m}\mathrm{Re}\big(\Psi^*\nabla\Psi\big)\cdot\nabla\right)\phi \\ &\quad - \int_{\mathcal{D}} \mathrm{d}S\,\frac{\hbar^2}{2m}\,\vec{n}\cdot\mathrm{Re}\big(\Psi^*\nabla\Psi\big)\restriction_{\mathcal{D}}\,\phi\restriction_{\mathcal{D}},\end{aligned} \tag{A3a}$$

$$
\begin{aligned}
0 = &\int_{\Omega} \mathrm{d}^3 r \left( \frac{\partial |\Psi|^2}{\partial t} - \frac{\hbar}{m} \mathrm{Im}\big(\Psi^* \nabla \Psi\big) \cdot \nabla \right) \xi \\
&+ \int_{\mathcal{D}} \mathrm{dS} \, \frac{\hbar}{m} \, \vec{n} \cdot \mathrm{Im}\big(\Psi^* \nabla \Psi\big) \restriction_{\mathcal{D}} \; \theta\big(+\vec{n} \cdot \mathrm{Im}(\Psi^* \nabla \Psi) \restriction_{\mathcal{D}}\big) \xi \restriction_{\mathcal{D}}
\end{aligned}
\tag{A3b}
$$

hold almost everywhere on $[0, \infty)$.